# The astrometric signal of microlensing events caused by free floating planets

**Lindita Hamolli[1], Mimoza Hafizi[1], Francesco De Paolis[2], Achille A. Nucita[2]**

[1] *Department of Physics, University of Tirana, Albania*

[2] *Department of Mathematics and Physics "Ennio De Giorgi'" and INFN, University of Salento, CP 193, I-73100 Lecce, Italy*

*e-mail: lindita.hamolli@fshn.edu.al*

**Abstract**

Astrometric observations of microlensing events can be used to obtain important information about lenses. During these events, the shift of the position of the multiple image centroid with respect to the source star location can be measured. This effect, which is expected to occur on scales from micro-arcseconds to milli-arcseconds, depends on the lens-source-observer system physical parameters. Here, we consider the astrometric and photometric observations by space and ground-based telescopes of microlensing events towards the Galactic bulge caused by free floating planets (FFPs). We show that the efficiency of astrometric signal on photometrically detected microlensing events tends to increase for higher FFP masses in our Galaxy. In addition, we estimate that during five years of the Gaia observations, about a dozen of microlensing events caused by FFPs are expected to be detectable.

## 1. Introduction

Gravitational microlensing is a well known technique for detecting compact objects in the disk, bulge and halo of our Galaxy via the amplification of the light of background sources. In the simplest case, when the point-like approximation for both lens and source is assumed, individual images cannot be resolved due to their small separation, but the total brightness of the images is larger with respect to that of the unlensed source, leading to a specific time dependent amplification of the source (Paczyński 1986, Paczyński 1996). This photometric effect has been used extensively in the search for extrasolar planets. In some cases, second order microlensing effects, as the finite source effect (see Refs. Nemiroff 1994, Witt & Mao 1994, Witt 1995, Hamolli et al. 2015) and the parallax effect (see Refs. Alcock 1995, Dominik 1998, Hamolli et al. 2013), may be extremely useful to solve the parameter degeneracy problem and strongly constrain the lens mass. In particular, it is used for free floating planets (FFPs), which are dark objects with mass $M \leq 0.01 M_e$ that are not bound to a host star (Sumi et al. 2011). Even though the microlensing events caused by FFPs are very short, the current microlensing syrveys are capable to detect them down to Earth-mass objects (Mróz et al. 2018). A consistent boost in the number of detectable FFPs is expected to occur by using space-based microlensing observations towards the Galactic bulge by, e. g. WFIRST telescope (Spergel et al. 2015, Strigari 2012).

In addition to the photometric effect, a gravitational microlensing event gives rise also to an astrometric deflection, as the event unfolds. This is because the images produced by the lens are not symmetrically distributed, leading to a typical elliptic pattern traced by the centroid, which was studied by many authors (Walker 1995, Miyamoto 1995, Hog 1995, Dominik & Sahu 2000, Nucita 2017). The ellipse semi-axes depend on the lens impact parameter $u_0$ and the Einstein angular radius, $\theta_E$ (see Sec. 2.2). The last one, for FFPs is of the order of µas (see Sec. 3 for more details). For this reason, the astrometric signal is expected to be detected more efficiently through space-based observations and indeed the Gaia satellite is the best instrument for such a mission. Gaia is a space observatory of the European Space Agency (ESA) launched on 19 December 2013 and is performing during 5 years the astrometric, photometric and spectroscopic observations. The precision of astrometric observations depends on the visual magnitude of the star. Eric Hog (2017) has determined the astrometric precision of the Gaia telescope and found that it can be as low as 4 µas for stars with visual magnitude in range from 6 to 13 (see table A in Ref. Hog 2017). While Gaia satellite is performing a survey of all the sky, follow-up astrometric observations of the interesting microlensing events are planned to be obtained by ground-based instruments as GRAVITY, which is a Very Large Telescope Interferometer build by the European Southern Observatory and Max Planck Institute for Extraterrestrial Physics. The interferometer will be used for precise astrometrical observations at a level around 10 µas (Eisenhauer et al. 2009, Zurlo et al. 2014).

Using Gaia as an astrometric microlensing survey instrument, we aim to determine the astrometric signal detected by it in microlensing events caused by FFPs toward galactic bulge, which in combination with present photometric measurements by ground-based telescopes (as OGLE) can be used to constrain the FFP masses and then their distances.

In fact, FFPs are dark objects, which may have been formed in protoplanetary disks at much smaller separations and then been scattered into unbound or very distant orbits. Since these objects are not surrounded by the plasma as in the case of black holes, we have not considered the plasma effect in the gravitational microlensing events caused by them (Bisnovatyi-Kogan 2012, Er & Mao, 2013).

The aim of the paper is to discuss the main observational features of the photometric and astrometric effects on the microlensing events caused by FFPs.

This paper is organized as follows. We review the amplification and the centroid shift of microlensing events in Sec. 2. In Sec. 3 we present the possibility of astrometric and photometric detections of microlensing events caused by FFPs. In Sec. 4 we show how the parameters of the microlensing events are simulated using Monte Carlo method and present our results for the detection of astrometric signal. In Section 5 we summarize our main conclusions.

## 2. Basics of photometric and astrometric microlensing

*2.1 Amplification of a microlensing event*

A microlensing event happens when the lens moves close enough to the line of sight towards the source star. In the simplest case, when the lens and the source are assumed to be point objects, there are two images, whose positions θ with respect to the lens are given by solving the lens equation (Schneider et al. 1992)

$$\underline{u}^2 - u\underline{u} - 1 = 0, \qquad (1)$$

where $u = \dfrac{\theta_S}{\theta_E}$ is the dimensionless distance of the source ($\theta_S$ - the angular distance of the source from the lens with mass M), $\underline{u} = \dfrac{\theta}{\theta_E}$ its images in the Einstein angular radius, $\theta_E$, which is given by

$$\theta_E = \sqrt{\frac{4GM}{c^2} \frac{D_S - D_L}{D_L D_S}} \approx 2851 \mu as \sqrt{\frac{M(M_e)}{D_L(kpc)}(1 - \frac{D_L}{D_S})} \; . \qquad (2)$$

Here, $D_S$ and $D_L$ are the distances to the source and the lens, respectively. The solutions of the second degree equation with respect to $\underline{u}$ are

$$\underline{u}_{+,-} = \frac{1}{2}\left(u \pm \sqrt{4 + u^2}\right), \qquad (3)$$

which give the locations of the positive and negative parity images (+ and −, respectively) with respect to the lens position. The magnification for each image is

$$A_{+,-} = \frac{1}{2}\left(1 \pm \frac{2 + u^2}{u\sqrt{4 + u^2}}\right), \qquad (4)$$

and the total magnification is given by (Paczyński 1986):

$$A = |A_+| + |A_-| = \frac{2 + u^2}{u\sqrt{4 + u^2}} \geq 1 . \qquad (5)$$

Note that, the total magnification depends only on the dimensionless source-lens distance, $u$. In the Figure 1 (left panel) are shown the positions of two images in the lens plane. The + image resides always outside a circular ring centered on the lens position with Einstein radius equal to the $R_E = \theta_E D_L$, while the − image is always within this ring. As the source-lens distance increases, the + image approaches to the source position (the blue line) while the − one (the magenta line) moves towards the lens location. Since $u$ varies with time, the equation (5) gives the source amplification as a function of time, i.e. the so-called Paczyński light curve.

## 2.2 The centroid shift of a microlensing event

During a microlensing event, the elliptic trajectory traced by the light centroid can be detected as well. To find its semi-axes, let us consider a source moving in the lens plane with transverse velocity $v_\perp$ and let be ξLη a frame of reference centered on the lens, with ξ and η axis oriented respectively along the velocity vector and perpendicular to it. The projected coordinates of the source (in units of the Einstein radius) are $\xi(t)=(t-t_0)/T_E$, $\eta(t)=u_0$ and the source-lens distance is $u^2=\xi^2+\eta^2$. Here $T_E=R_E/v_\perp$ is the Einstein time scale of the event and $u_0$ is the minimum projected distance of the lens to the source (in this case lying on the η axis) at time $t_0$. As the event unfolds, the two images are in the same line with the lens and the source, but in the opposite side. The centroid of the image pair, which is defined as the average position of + and − images weighted by the associated magnifications (Walker 1995, Boden et al.1998, Nucita et al. 2016), is given by (its position related to the lens)

$$\bar{u} = \frac{u_+ A_+ + u_- A_-}{A_+ + A_-} = \frac{u(u^2+3)}{u^2+2}. \tag{6}$$

It is always in the same line with the lens and the source, but on the side of the + image.

The measurable quantity during an event is the displacement Δ of the centroid of the image pair relative to the source, i.e.

$$\Delta = \bar{u} - u = \frac{u}{u^2+2} = \frac{\sqrt{(\frac{t-t_0}{T_E})^2 + u_0^2}}{(\frac{t-t_0}{T_E})^2 + u_0^2 + 2} = f(t, u_0), \tag{7}$$

which is a function of $t$ and $u_0$.

Since, in the frame ξLη $u$ is a vector, then Δ can also be seen as a vector with components ($\Delta_\xi, \Delta_\eta$) along the axes ξ and η

$$\Delta_\xi = \frac{\xi(t)}{u^2+2}, \quad \Delta_\eta = \frac{\eta(t)}{u^2+2}. \tag{8}$$

In Figure 1 (the right panel) are shown the positions of the two images and the centroid shift related to the source.

The components of Δ vary with time, where $\Delta_\eta$ is symmetric with respect to $t_0$ and always positive, while $\Delta_\xi$ is an antisymmetric function with minimum and maximum values occurring at $t=t_0 \pm T_E\sqrt{u_0^2+2}$, respectively (see Figure 2).

One can also verify that the maximum centroid shift is $\sqrt{2}/4 \approx 0.35$, for $u=\sqrt{2}$ (see equation 7). For microlensing events with impact parameters $u_0 < \sqrt{2}$, the centroid shift goes through two maxima at $t=t_0 \pm T_E\sqrt{u_0^2+2}$ and for $u=\sqrt{2}$ it tends linearly to zero. But, the opposite happens with the photometric amplification, which increases towards small values of $u$. In addition, for large values of $u$, the centroid shift falls slower than the magnification, which could be a promising observable for large source-lens distances, i.e. relatively far from the light curve peak (see Figure 3).

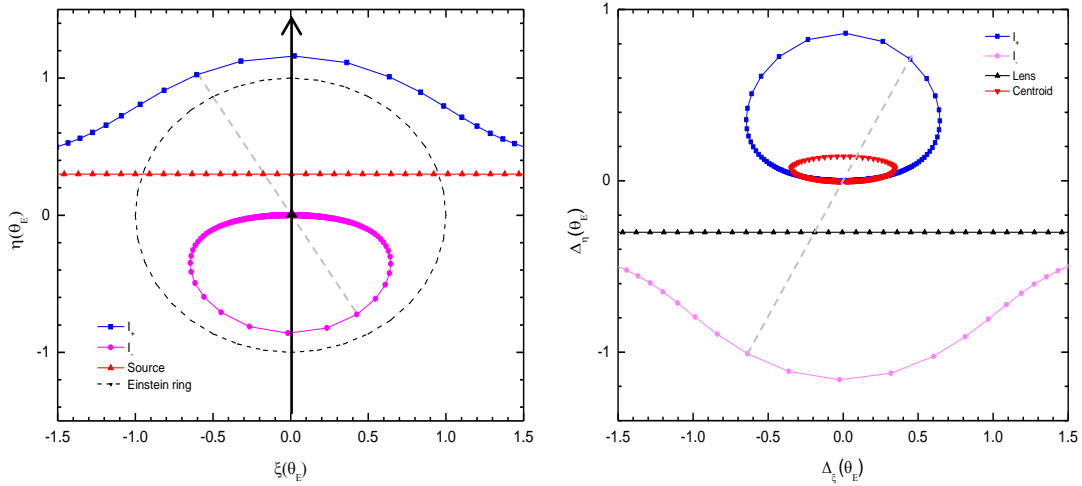

**Figure 1**. The positions of the two images ($I_+$, $I_-$) of a star ($D_S = 8.5 kpc$) during a microlensing event caused by a lens ($M = 10^{-3} M_e$, $D_L = D_S/2$, $u_0 = 0.3$, $v_\perp = 100\ km/s$) in the plane $\xi,\eta$ centered on the lens (leftpanel) and in the plane $\Delta\xi, \Delta\eta$ centered on the star (right panel). The black dashed line shows the Einstein ring.

In microlensing events with $u_0 \geq \sqrt{2}$, $\Delta$ gets the single maximum value equal to $u_0/(u_0^2 + 2)$ at $t = t_0$.

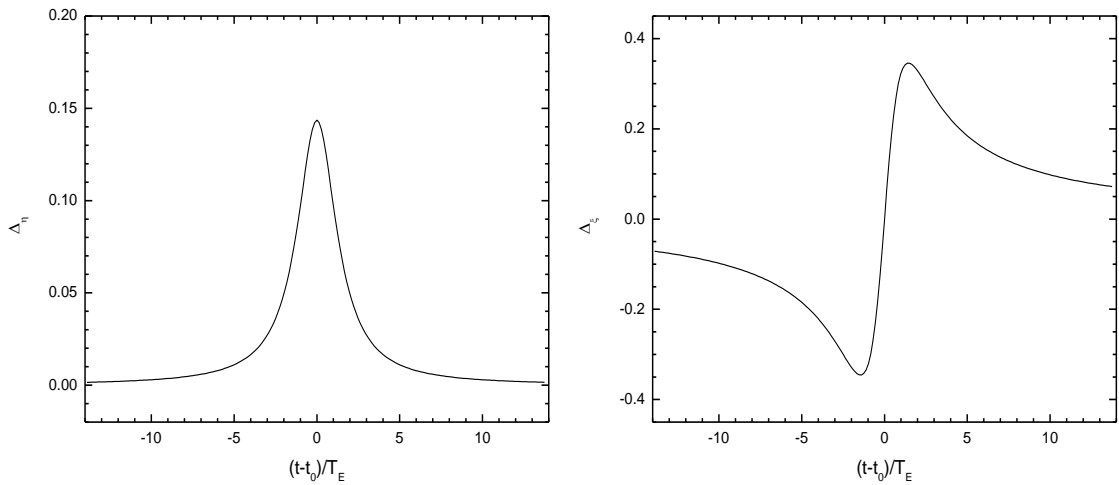

**Figure 2.** For a microlensing event with $u_0 = 0.3 < \sqrt{2}$ we show $\eta$ (left panel) and $\xi$ (right panel) components of the centroid shift as a function of $(t-t_0)/T_E$.

Unlike A, which is a dimensionless quantity that depends only on the dimensionless separation $u$, $\Delta$ is a function of both $u$ and $\theta_E$, such that for a given $u$, the observed centroid shift is directly proportional to $\theta_E$ (Dominik & Sahu 2000),

$$\Delta = \frac{u}{u^2 + 2} \theta_E . \qquad (9)$$

In Figure 4 (left panel) we have plotted the ellipse that traces the light centroid for a specific case. The ellipse semi-axes a (along $\Delta_\xi$) and b (along $\Delta_\eta$), which depend on both the lens impact parameter $u_0$ and $\theta_E$, are given by

$$a = \frac{1}{2} \frac{\theta_E}{\sqrt{u_0^2 + 2}} \;,\; b = \frac{1}{2} \frac{u_0}{u_0^2 + 2} \theta_E . \qquad (10)$$

It is evident that for $u_0 \to \infty$ the ellipse becomes a circle with radius $\theta_E/2u_0$ and for $u_0$ approaching zero it becomes a straight line of length $\theta_E/\sqrt{2}$ (see e.g., Walker 1995, Nucita et al. 2017).

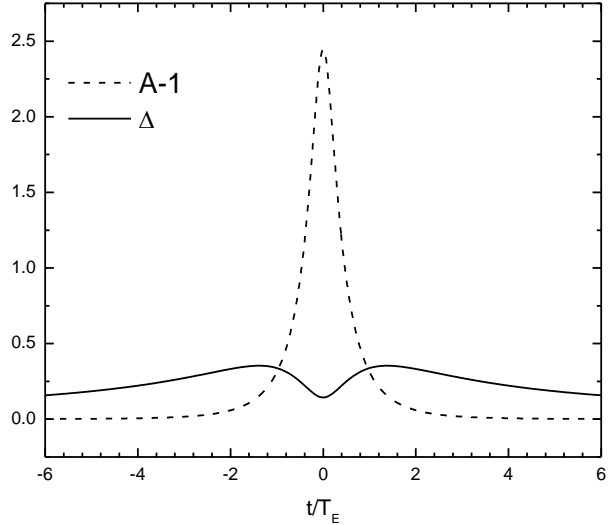

**Figure 3.** The amplification A-1 (dashed line from eq. 5) and the centroid shift (continuous line) of a microlensing event with $u_0 = 0.3 < \sqrt{2}$ are plotted as a function of $(t-t_0)/T_E$. The source amplification is greatest at the minimum separation between the lens and the source, whereas the astrometric shift reaches two maxima at $(t-t_0)/T_E = \pm\sqrt{u_0^2 + 2}$ and a minimum at $(t-t_0)/T_E = 0$. We remark that $\Delta$ falls slower than the photometric amplification for large lens-star separation.

We assume that a microlensing event can be detected astrometrically if the centroid shift is larger than the astrometric threshold of the telescope $\delta_{th}$. For $u_0 \leq \sqrt{2}$, the maximum value of the centroid shift $\Delta_{max} \simeq 0.35\theta_E$ depends only on $\theta_E$, while for $u_0 > \sqrt{2}$ it also depends on the value of $u_0$, since $\Delta_{max} = [u_0/(u_0^2 + 2)]\theta_E$.

In Figure 4 (right panel) we show a microlensing event with centroid shift larger than the astrometric threshold, $\delta_{th}$. It is therefore an event astrometrically detectable. The $\delta_{th}$ value is given in units of $\theta_E$. It is evident that when a microlensing event is astrometrically detectable, the lens parameters can be better constrained with respect to the case when the event is detectable only photometrically. A further advantage of the astrometric microlensing technique is that an event is potentially observable for a much longer time with respect to the typical photometric event, because astrometric signals persist much longer (i.e. for larger lens-source separations) than photometric signals.

The importance of astrometric microlensing observations is that one can determine the values of both $\theta_E$ and $u_0$ from the observed astrometric ellipse. This is because the size (semi-major and semi-minor axes) of the astrometric ellipse is directly proportional to $\theta_E$ (see equation 10). Moreover, for events with $u_0 \leq \sqrt{2}$, the Einstein time $T_E$ can be readily derived by measuring the time lag between the peak features.

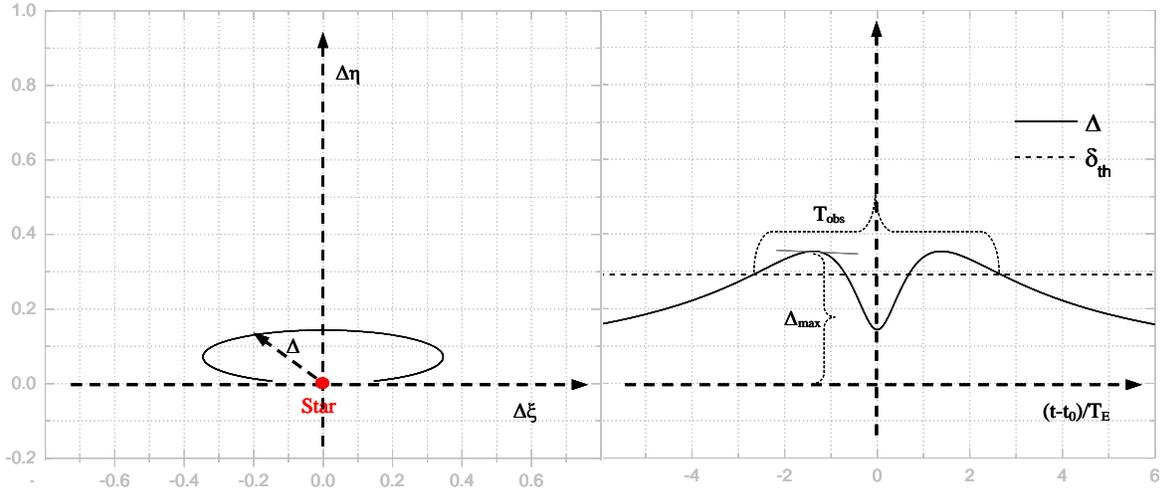

**Figure 4**. Left panel: The centroid shift ellipse of the source star during a microlensing event with parameters: $D_S = 8.5 kpc$, $M = 10^{-3} M_e$, $D_L = D_S/2$, $u_0 = 0.3$ and $v_\perp = 100\ km/s$ is shown in the $\Delta\xi$, $\Delta\eta$ plane. Right panel: The variation of the centroid shift (continuous line) as a function of $(t-t_0)/T_E$ ($t_0$ is considered to be zero). It goes through two maxima at $t_0 \pm T_E \sqrt{u_0^2 + 2}$ equal to $\sqrt{2}/4$; 0.35 and a minimum at $t = 0$, with value equal to $2b = 0.15$. In this case, the centroid shift traces an ellipse with semi-major axis $a = 0.35$. The dashed line shows the astrometric threshold. Here all the angular quantities are given in units of the Einstein angle $\theta_E$.

In the case of the usual photometric microlensing, a commonly used signature is the amplification of the source light with respect to the threshold value $A_{th}$ at a given time which, for ground-based observations, is considered to be 1.34. The photometric observation time, $T_{phot}$ can be defined by using the equations $u_{th} = \sqrt{(2/\sqrt{1-A_{th}^{-2}})-2}$ and $u_{th} = \sqrt{(T_{phot}/2-t_0)^2/T_E + u_0^2}$.

In the same way, we define the astrometric observation time, $T_{ast}$, as the time during which the event is expected to be detectable astrometrically. To this aim, we require that the value of the centroid shift has to exceed a given threshold $\delta_{th}$. We use the criterion $\Delta(t) = \frac{u(t)}{u(t)^2+2}\theta_E \geq \delta_{th}$ to calculate the astrometric observation time. Depending on the value of $u_0$, the above inequality has four solutions when $u_0 < \sqrt{2}$ (see the Figure 4) and two when $u_0 > \sqrt{2}$. If the $\delta_{th}$ line does not intersect the variation curve of the centroid shift, this means that $\Delta_{max} < \delta_{th}$ and the above mentioned inequality has no solution. But, since during an event only Δ(t) can be measured, which is the variation of the source star centroid shift with the time, we also consider the change of the centroid shift between two points of the ellipse, called variable centroid shift $\Delta_{var}$. If $\Delta_{var}$ is larger than $\delta_{th}$, this microlensing event can be detected astrometrically and we can find the time during which the variable centroid shift is larger than the threshold value, $\delta_{th}$. Note that as an astrometric observation time, $T_{ast}$ we consider the largest value obtained by $\Delta(t) \geq \delta_{th}$ and $\Delta_{var} \geq \delta_{th}$.

## 3. Astrometric microlensing due to free floating planets

The microlensig events caused by FFPs are generally photometrically short events, because their Einstein angular radius is very small. The FFP mass is considered to be in the range $[10^{-5}, 10^{-2}]M_e$. If we consider a source star in the bulge of our Galaxy ($D_S = 8.5$ kpc) and the lens in the middle of the observer-source distance, the Einstein angular radius results to be in the range $3 \div 98$ $\mu as$ (see equation 2). So, for microlensing events with $u_0 \leq \sqrt{2}$, the maximum value of the centroid shift ($\Delta_{max}$; $0.35\theta_E$) will be in the range $1 \div 35$ µas. These events are astrometrically detectable if the precision of the astrometric observation is good enough, i. e. if the requirement $\Delta_{max} \geq \delta_{th}$ or $\Delta_{var} \geq \delta_{th}$ is satisfied.

Our aim in this paper is to investigate the astrometric signal caused by FFPs which, in combination with photometric measurements can be used to precisely constrain the FFP mass. To estimate the mass of the lens $M = c^2 r_E \theta_E / 4G$, we need the value of both $r_E$ (the Einstein radius projected onto the observer plane), which can be estimated if the parallax effect in microlensing events is detected, and $\theta_E$, which can be measured by the finite source effect. In a previous analysis (Hamolli et al. 2015) we have found that for about 1/3 of all microlensing events observed by space-based telescopes (as Euclid) and caused by FFPs is possible to have detectable finite source effects. Another possibility for measuring $\theta_E$ is the astrometric observation.

In this work, we calculate the efficiency of the astrometric signal detection in microlensing events caused by the FFPs toward Galactic bulge. To this aim, we consider in particular ground-based observations as those by OGLE telescope and astrometric observations by Gaia. Since FFP microlensing events are short time events, we have not considered the parallax effect despite the rotation of the Earth and Gaia satellite around the Sun. During the 2016 campaign, the OGLE telescope has observed 1927 microlensing events towards the Galactic bulge. The visual magnitude of the stars varies from 12 to 21, with an observational sampling of the images ranging from minutes to days. For Gaia observations, Eric Hog (2017) has defined the median astrometric accuracy as function of the star magnitude. His results are given in Table 1.

Table 1. Median astrometric precision in µas for Gaia observations as function of the source star apparent magnitude.

| Star magnitude | 6<m<13 | 13<m<14 | 14<m<15 | 15<m<16 | 16<m<17 | 17<m<18 | 18<m<19 | 19<m<20 |
|---|---|---|---|---|---|---|---|---|
| Precision (µas) | 4 | 6 | 9 | 13 | 21 | 34 | 56 | 98 |

## 4. Monte Carlo calculations and results

A microlensing event is considered detectable if there are at least eight consecutive points with amplification larger than the threshold amplification. We take into account the microlensing events as those observed by OGLE collaboration and assume that the threshold amplification value is the minimum value of the peak amplification of the OGLE events detected during the 2016 campaign, that is $A_{th} = 1.027$. This value corresponds to the impact parameter smaller than $u_{0\_phot} = 2.6$. Concerning the astrometric observations we consider the Gaia telescope for which the astrometric precision depends on the magnitude of the source stars, as is shown above. We assume that a microlensing event is astrometrically detected if $\Delta_{max} \geq \delta_{th}$ or $\Delta_{var} \geq \delta_{th}$. For a lens with mass $M_L = 10^{-2} \, M_e$ (the upper limit of the FFP mass range) in the middle of the observer-source distance ($D_L = 4.25$ kpc) and $\delta_{th} = 4 \mu as$ (the smallest value of the astrometric precision) we find that the maximum value of the impact parameter for a microlensing event astrometrically detectable is equal to $u_{0\_ast} = 13.6$

To investigate the possibility to measure the astrometric and photometric signals in microlensing events caused by the FFPs in the thin disk towards the Galactic bulge, we use Monte Carlo numerical simulations.

For each microlensing event we simulate five parameters: the FFP distance, mass, transverse velocity, impact parameter and the magnitude of the source star, while its distance is considered fixed at 8.5 kpc. In particular:

1) We generate the FFP distance from the observer $D_L$, based on the thin disk FFP spatial distributions, which is assumed to follow that of the stars (Gilmore et al. 1989, De Paolis et al. 2001, Hafizi et al. 2004). The galactic coordinates of the line of sight toward the Galactic bulge are considered ($l = 1.1^o$, $b = -1.6^o$).

2) We extract the lens mass M following the FFP mass function $dN/dM : k_{PL} M^{-\alpha_{PL}}$ defined by Sumi et al. (2011), where the mass function index $\alpha_{PL}$ takes values in the range [0.9; 1.6].

3) We generate the FFP relative transverse velocity assuming for each component the Maxwellian distribution

$$f(v_i) \propto e^{-\frac{(v_i - \bar{v}_i)^2}{2\sigma_i^2}}, \quad i \in \{x, y, z\} \tag{11}$$

(see Refs. Han & Gould 1995 and Hamolli et al. 2016 for details).

4) The event impact parameter $u_0$ is randomly extracted from a uniform distribution in the interval [0, 13.6]. The upper value of this interval is considered the highest value between $u_{0\_phot} = 2.6$ and $u_{0\_ast} = 13.6$.

5) The apparent magnitude of the star is defined by using the star mass function

$$dN : \begin{cases} M^{-2.2} dM \, (0.5 \leq \frac{M}{M_e} < 1) \\ M^{-2.7} dM \, (1 \leq \frac{M}{M_e} < 100) \end{cases} \tag{12}$$

and the standard mass-luminosity relation $L/L_e = (M/M_e)^\alpha$ (with $\alpha$ ; 3.5). Since Gaia telescope observes stars of apparent magnitude up to 20, we exclude the stars with apparent magnitude bigger than 20, retaining only the stars with magnitude that, during the microlensing event becomes, lower than 20. The absolute magnitude of the Sun is considered to be equal to 4.83. The astrometric precision assumed for Gaia observations as function of the source star apparent magnitude is given in Table 1.

We simulate 1000 microlensing events and for each of them calculate the photometric light curve, the centroid shift and the variable centroid shift. Based on the above conditions, we define if the event is detected photometrically, astrometrically or both simultaneously. Since the OGLE observations have not a fixed cadence, we consider three values for it: 30 min, 2 hours and 1 day.

The astrometric efficiency of the events photometrically detectable is calculated as the ratio of the number of events with detectable astrometric and photometric signal to the number of events photometrically detectable. The photometric efficiency of the events astrometrically detectable is calculated as the ratio of the number of events with detectable astrometric and photometric signal to the number of events astrometrically detectable.

We have considered the FFP distribution in the thin disk and defined the efficiencies with respect to the value of $\alpha_{PL}$ in the range 0.9 -1.6.

As one can see, the astrometric efficiency of the events photometrically detectable depends on the index value of FFP mass function. Since the centroid shift of source star images falls slower than its magnication, the number of events astrometrically detectable does not depend substantially on the observational cadence. But, the number of microlensing events photometrically detectable decreases when the cadence of observations is increased. So, when the cadence of observation increases, the astrometric efficiency of the events photometrically detectable is increased.

Table 2. The astrometric and photometric efficiencies for microlensing events caused by FFPs in the thin disk for cadences: 30 min, 2 hour and 1 day. The α$_{PL}$ value is considered in range from 0.9 to 1.6.

| α$_{PL}$ | Cadence 30 min | | Cadence 2 hours | | Cadence 1 day | |
|---|---|---|---|---|---|---|
| | Astrometric efficiency | Photometric efficiency | Astrometric efficiency | Photometric efficiency | Astrometric efficiency | Photometric efficiency |
| 0.9 | 0.116 | 0.606 | 0.119 | 0.606 | 0.233 | 0.515 |
| 1.0 | 0.099 | 0.630 | 0.104 | 0.630 | 0.276 | 0.615 |
| 1.1 | 0.076 | 0.591 | 0.080 | 0.591 | 0.244 | 0.500 |
| 1.2 | 0.070 | 0.600 | 0.075 | 0.600 | 0.294 | 0.500 |
| 1.3 | 0.064 | 0.611 | 0.071 | 0.611 | 0.310 | 0.500 |
| 1.4 | 0.053 | 0.692 | 0.059 | 0.692 | 0.308 | 0.667 |
| 1.5 | 0.041 | 0.700 | 0.047 | 0.700 | 0.222 | 0.400 |
| 1.6 | 0.036 | 0.750 | 0.041 | 0.750 | 0.250 | 0.375 |

A significant point to consider is how the astrometric signal varies with respect to $D_L$. To answer this question we simulated 1000 microlensing events for each bin of $D_L$ defined in range [0, 1] ÷ [7, 8] kpc. Considering $\alpha_{PL} = 1.3$ and cadence = 2 hours (the middle value) we define the number of events detected astrometrically and then how many of them are detected photometrically (see Table 3). Likewise, we define the number of microlensing events detected photometrically and then how many of them are detected astrometrically (see Table 4).

In Figure 5 we have shown the efficiency of events astrometrically detectable (the number of events astrometrically detectable divided by the number of generated events), the efficiency of events photometrically detectable (the number of events photometrically detectable divided by the number of generated events) and the efficiency of events detectable both astrometrically and photometrically (the number of events detectable simultaneously in both ways divided by the number of generated events).

As one can see, the astrometric signal efficiency decreases when the value of $D_L$ is increased.

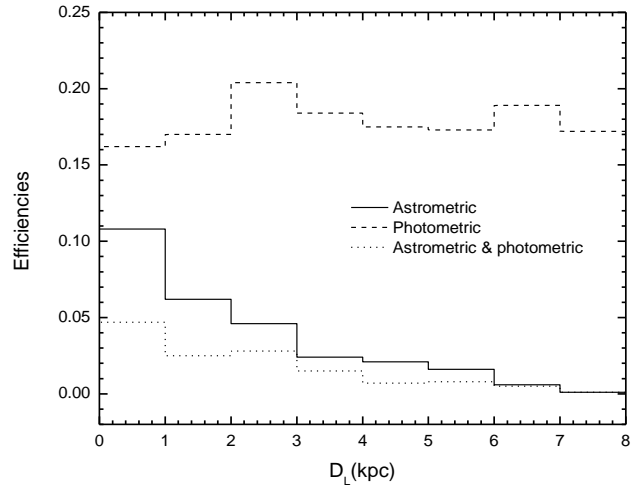

**Figure 5.** The efficiencies of FFP microlensing events towards the Galactic bulge are given as a function of the lens distance $D_L$ (in kpc). The continuous line shows the efficiency of events detectable astrometrically, the dashed line shows efficiency of events detectable photometrically and the dotted line shows the efficiency of events detectable both astrometrically and photometrically. Here the FFPs are assumed to be in the Galactic disk.

**Table 3**. The expected number of microlensing events astrometrically detectable for 1000 generated events with the procedure indicated in the text is given in the first row for 8 bins in $D_L$ (in kpc). The second row gives the fraction of the events given in the first row which are photometrically detectable.

|  | 0<$D_L$<1 | 1<$D_L$<2 | 2<$D_L$<3 | 3<$D_L$<4 | 4<$D_L$<5 | 5<$D_L$<6 | 6<$D_L$<7 | 7<$D_L$<8 |
|---|---|---|---|---|---|---|---|---|
| Number of events astrometrically detectable | 108 | 62 | 46 | 24 | 21 | 16 | 6 | 1 |
| Number of astrometric events photometrically detectable | 47 | 25 | 28 | 15 | 7 | 8 | 5 | 1 |

**Table 4**. The expected number of microlensing events photometrically detectable for 1000 generated events with the procedure indicated in the text is given in the first row for 8 bins in $D_L$ (in kpc). The second row gives the fraction of the events given in the first row which are astrometrically detectable.

|  | 0<$D_L$<1 | 1<$D_L$<2 | 2<$D_L$<3 | 3<$D_L$<4 | 4<$D_L$<5 | 5<$D_L$<6 | 6<$D_L$<7 | 7<$D_L$<8 |
|---|---|---|---|---|---|---|---|---|
| Number of events photometrically detectable | 162 | 170 | 204 | 184 | 175 | 173 | 189 | 172 |
| Number of photometric events astrometrically detectable | 47 | 25 | 28 | 15 | 7 | 8 | 5 | 1 |

## 5. Conclusions

In this paper we have considered the astrometric and photometric signal of microlensing events caused by the FFPs towards the Galactic bulge taking into account in particular the observing capabilities of the GAIA and OGLE telescopes, respectively.

Using Monte Carlo numerical simulations, we find that the efficiency of astrometric signal on microlensing events photometrically detectable tends to get smaller with increasing the index value of FFP mass function. At the same time, when the cadence of observation increases, the efficiency of astrometric signal on microlensing events photometrically detectable tends to get larger.

Moreover, we find that the detection efficiency of the astrometric effect varies from about 10%, for close FFPs, to zero for FFPs in the Galactic bulge. On the contrary, the photometric detection efficiency is roughly independent from $D_L$ and turns out to be between 15% and 20%. We also find that the detection efficiency for events astrometrically and photometrically detectable is only a few percent for FFPs belonging to the Galactic disk.

If the microlensing events with $T_E$ smaller than 2 days are considered as events caused by FFPs (Sumi et al. 2011), there are 34 of them detected by OGLE observations during 2016[1]. For $\alpha_{PL} = 1.3$ and cadence 2 hours, we find that 2÷3 of them could have astrometric signal observed by Gaia. So, during five years of the Gaia observations, we could estimate that about a dozen of FFP microlensing events may be detected both astrometrically and photometrically. These events would be able to constrain the FFP mass and distance distribution, which are fundamental information for FFP populations throughout the Milky Way.

*Acknowledgements.* FDP and AN acknowledge the support by the INFN projects TAsP and Euclid, respectively.

  

---

[1] http://ogle.astrouw.edu.pl/ogle4/ews/ews.html